\documentclass[twoside,slac_one]{revtex4}
\usepackage{graphicx}
\usepackage{fancyhdr}
\usepackage{amsmath} 
\usepackage{bm}
\usepackage{amsxtra}
\usepackage{amssymb}
\usepackage{amsthm}
\usepackage{latexsym}
\usepackage{lscape}

\def\t13{\theta_{13}}
\def\st13{\sin^2 2\theta_{13}}
\def\anue{\bar\nu_e}

\newcommand{\si}[2]{\sin\theta_{#1#2}}
\newcommand{\co}[2]{\cos\theta_{#1#2}}

\begin{document}

\title{Daya Bay Neutrino Experiment: Goal, Progress and Schedule}

%

\author{Zhe Wang}
\thanks{On behalf of the Daya Bay Collaboration}
\affiliation{Physics Department, Brookhaven National Laboratory, USA}

\begin{abstract}
Daya Bay Neutrino Experiment is dedicated to measuring the last unobserved neutrino mixing angle $\t13$.
The predicted precision on $\st13$ is 0.01 at 90\% confidence level. This document briefly reviews the
measurement method and detector construction status. The first two anti-neutrino detectors'
dry run result is also discussed. The Daya Bay near hall data taking is expected to commence in the summer of 2011
and the data taking of all of the three halls in the summer of 2012.
\end{abstract}

\maketitle

\thispagestyle{fancy}


\section{Introduction}
Based on a three-generation assumption, a $3\times3$ mixing matrix was proposed to
explain the oscillation in neutrino propagation, which is usually called
Pontecorvo-Maki-Nakagawa-Sakata matrix (PMNS matrix).
Eq.~\ref{eq:PMNS} shows one possible parametrization of the PMNS matrix which, for
dirac neutrino assumption, includes three mixing angles, $\theta_{12}$, $\theta_{23}$, $\theta_{13}$
and one CP-violation phase angle, $\delta$.
\begin{equation}
PMNS=
\left[\begin{array}{ccc}
1 & 0          & 0          \\
0 & \co{2}{3}  & \si{2}{3}  \\
0 & -\si{2}{3} & \co{2}{3}
\end{array} \right]
\left[\begin{array}{ccc}
\co{1}{3}  & 0 & \si{1}{3}e^{-i\delta} \\
0          & 1 & 0         \\
-\si{1}{3}e^{i\delta} & 0 & \co{1}{3}
\end{array} \right]
\left[\begin{array}{ccc}
\co{1}{2}  & \si{1}{2} & 0 \\
-\si{1}{2} & \co{1}{2} & 0 \\
0     & 0    & 1
\end{array} \right]
\label{eq:PMNS}
\end{equation}
Like the mixing matrix to describe quark mixing, these parameters cannot be
predicted by theory, and their values must be determined by experimental measurement.
A global fit~\cite{global_fit} and citations therein show $\theta_{12}$ is close to 34$^o$ and $\theta_{23}$ is close to 45$^o$.
In the past Chooz~\cite{Chooz} obtained the best limit of 0.17 in $\st13$
for $\Delta m^2_{31} = 2.5\times10^{-3} eV^2$ at the 90\% confidence level.
The particular pattern of the PMNS matrix, two big mixing angles and a small one,
is an intriguing picture, which might indicate some discrete family symmetries,
like the Tri-bimaximal mixing model~\cite{Tri-bimaximal}, etc.
Besides the interest of its value in theory, $\t13$ is also a critical input for
other neutrino experiments. As can be seen from Eq.~\ref{eq:PMNS}, the CP-violation
phase term, $e^{i\delta}$ is always multiplied by $\sin\t13$.
A zero $\t13$ will make any CP-violation undetectable.
$\t13$ also affects the sensitivity to the neutrino mass hierarchy in future long baseline neutrino
experiment~\cite{LongBaseline} . $\t13$ is also present in the effective electron neutrino mass term, which is
usually the direct measurement of neutrinoless double beta decay experiments.
To really understand the majorana mass of neutrino, a reasonable estimation of $\t13$ is always necessary.
Ultimately the unitarity of the PMNS matrix can be tested, i.e. the 3-flavor mixing framework, possibly
leading to further exciting discoveries.

The Daya Bay Neutrino Experiment~\cite{Dyb_arxiv} exploits reactor generated anti-electron-neutrinos, $\anue$, with
gadolinium (Gd) loaded liquid scintillator (LS), measures their disappearance probability,
then extracts the mixing angle, $\t13$. The Daya Bay Experiment plans to measure $\st13$ to
a precision of 0.01 at 90\% confidence level with three years' data taking.

In the second section the measurement method will be briefly reviewed.
The status of the detector construction is discussed in the third section.
In the fourth section the detector dry run results will be shown,
then a short summary and outlook are given in the last section.

\section{Measurement Method}
\subsection{Detection of anti-electron-neutrinos}
Low energy $\anue$, mostly less than 10MeV, are emitted by the fission of nuclei in reactors.
They can be detected by the inverse-beta-decay (IBD) process, i.e.
$\anue$ are captured by the free protons (hydrogen) in the target region of the Daya Bay detector.
\begin{equation}
\anue + p \rightarrow e^+ + n
\end{equation}
where the energy loss and annihilation of $e^+$ gives a prompt signal.
This process then is followed by the capture of the neutron which gives a delayed signal.
At Daya Bay two dominant types of captures can happen.
One type is capture on Gd:
\begin{equation}
n + Gd \rightarrow Gd^* \rightarrow Gd + \gamma's
\end{equation}
The capture process on Gd will release on average three or four photons, and their total energy
is about 8MeV. For liquid scintillator with 0.1\% Gd in mass, the average capture lifetime is 28$\mu s$.
The other type of capture is on hydrogen:
\begin{equation}
n + p \rightarrow D + \gamma
\end{equation}
which will release a 2.2 MeV single gamma. Its average capture lifetime is 180$\mu s$.

\subsection{Precise Measurement of $\st13$}
The concept of the measurement of $\st13$ at Daya Bay can be described by a very short equation of the form:
\begin{equation}
\frac{N_f}{N_n} = \left(\frac{N_{p,f}}{N_{p,n}}\right)
                  \left(\frac{L_n}{L_f}\right)^2
                  \left(\frac{\epsilon_f}{\epsilon_n}\right)
                  \left[\frac{P_{sur}(E,L_f)}{P_{sur}(E,L_n)} \right]
\end{equation}
where $N_f$ and $N_n$ are measured numbers of IBD event at far and near sites respectively.
Then $N_{p,f}$ and $N_{p,n}$ are the numbers of free protons,
$L_f$ and $L_n$ are the baselines, $\epsilon_f$ and $\epsilon_n$ the IBD event detection efficiencies
for far and near site respectively.
The $P_{sur}(E,L_f)$ indicates the survival probability of $\anue$ with energy, E, at baseline $L_f$.
Similarly $P_{sur}(E,L_n)$ is for near site.

Earlier attempts to measure $\st13$ usually had only one detector, measuring the
$\anue$ flux at some distance, then comparing that flux to a prediction
estimated based on the thermal power output of the reactor. The intrinsic uncertainty in the neutrino flux
prediction made on the basis of thermal power is around 2-3\% which is not sufficient for a precise
measurement of $\t13$. The Daya Bay Experiment has near detectors that are close to the reactor and
can absolutely calibrate the flux before oscillation.

The discrepancy between near and far detectors needs to be minimized.
The Daya Bay Experiment is trying to make detector modules for every site identical in all aspects
from mechanical design, manufacture and liquid scintillator synthesis.
Many systematic errors can be canceled in this way.
Detector swapping is also planned to crosscheck their performance.

The large statistics given by the powerful reactor and high target mass
is another precondition for making a precise measurement.

All the modules are located underground.
Cosmic ray muon flux is greatly suppressed and in addition an active muon veto detector is deployed to identify muon-induced background.

The Daya Bay Experiment plans to measure $\st13$ to
a precision of 0.01 at 90\% confidence level with three years' data taking.
This can be interpreted as follows: if the true value of $\st13$ is zero, the final 90\% coverage uncertainty of
it contributed by all statistical and systematic error sources is 0.01.
The real setup of the Daya Bay Experiment is a little more complicated than this
simple two-detector situation.
Details about the sensitivity prediction of the Daya Bay Neutrino Experiment can be
found in~\cite{Dyb_arxiv}.

\section{Detector Construction}
\subsection{Site}
The Daya Bay Nuclear Power Plant is a complex composed of three sites, all located along the coast line.
They are called ``Daya Bay'', ``Ling Ao I'' and ``Ling Ao II'' respectively.
Each site includes two reactor cores.
By the summer of 2011 all reactors were put into commercial running.
The total nominal thermal power is 6$\times$2.9 GW,
which is one of the most powerful nuclear power plants in the world.

The Daya Bay Neutrino Experiment has three experimental halls. Two near halls, "Daya Bay Near Hall" and
"Ling Ao Near Hall", are used to calibrate the $\anue$ flux from reactors. A far hall, "Far Hall",
is located at a distance which has maximum sensitivity for $\anue$ disappearance probability detection.
The baselines and overburdens of the three halls are tabulated in Tab.~\ref{tab:sites}.
The simulation-estimated muon fluxes and IBD event rates at three halls are also listed in Tab.~\ref{tab:sites}.

Each near hall has two anti-neutrino detector (AD) modules. They are immerged in
an octagonal water pool serving as active muon veto.
On the top of the water pool is a layer of resistive plate chambers, RPCs,
also serving as a muon veto and giving better position information.
The Far Hall has a similar design, but to increase statistics, four ADs are needed and correspondingly
larger water pool and RPC module arrays.

\begin{table}
\begin{center}
\caption{Target mass, baseline, overburden, estimated muon flux and IBD rate at three halls.}
\begin{tabular}{|l|c|c|c|}
\hline  & Daya Bay near hall & Ling Ao near hall &  Far hall \\
\hline Target mass & 40 tons & 40 tons & 80 tons \\
\hline Baseline    & about 360 m            & about 500 m                 & about 1600 m  \\
                   & (to Daya Bay reactors) & (to Ling Ao I, II reactors) & (to Daya Bay reactors) \\
                   & &  &                                                   about 1900 m \\
                   & &  &                                                   (to Ling Ao I, II reactors) \\
\hline Overburden  & 98 m   & 112 m  & 350 m \\
\hline Estimated Muon flux & 1.2 Hz/m$^2$ & 0.73 Hz/m$^2$ & 0.04 Hz/m$^2$ \\
\hline Estimated IBD rate  & 840/day/module      & 740/day/module       & 90/day/module \\
\hline
\end{tabular}
\label{tab:sites}
\end{center}
\end{table}

\subsection{Anti-neutrino Detector}
\begin{figure}
\centering
\includegraphics[width=80mm]{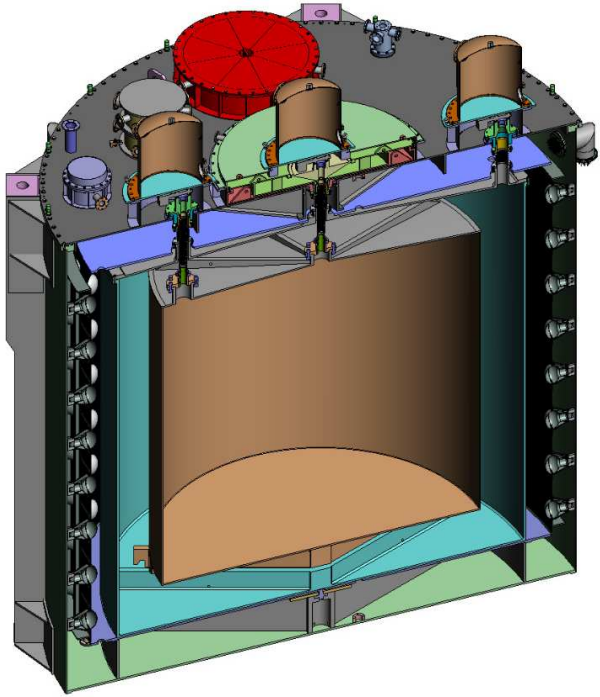}
\caption{AD sectional drawing. From inside to outside are the volumes for Gd-loaded liquid scintillator, liquid scintillator and mineral oil. Eight rings of PMTs are deployed along the outer stainless steel wall. On the top are the three ACU units.}
\label{fig:AD-sectioned}
\end{figure}

\begin{figure}
\centering
\includegraphics[width=120mm,height=90mm]{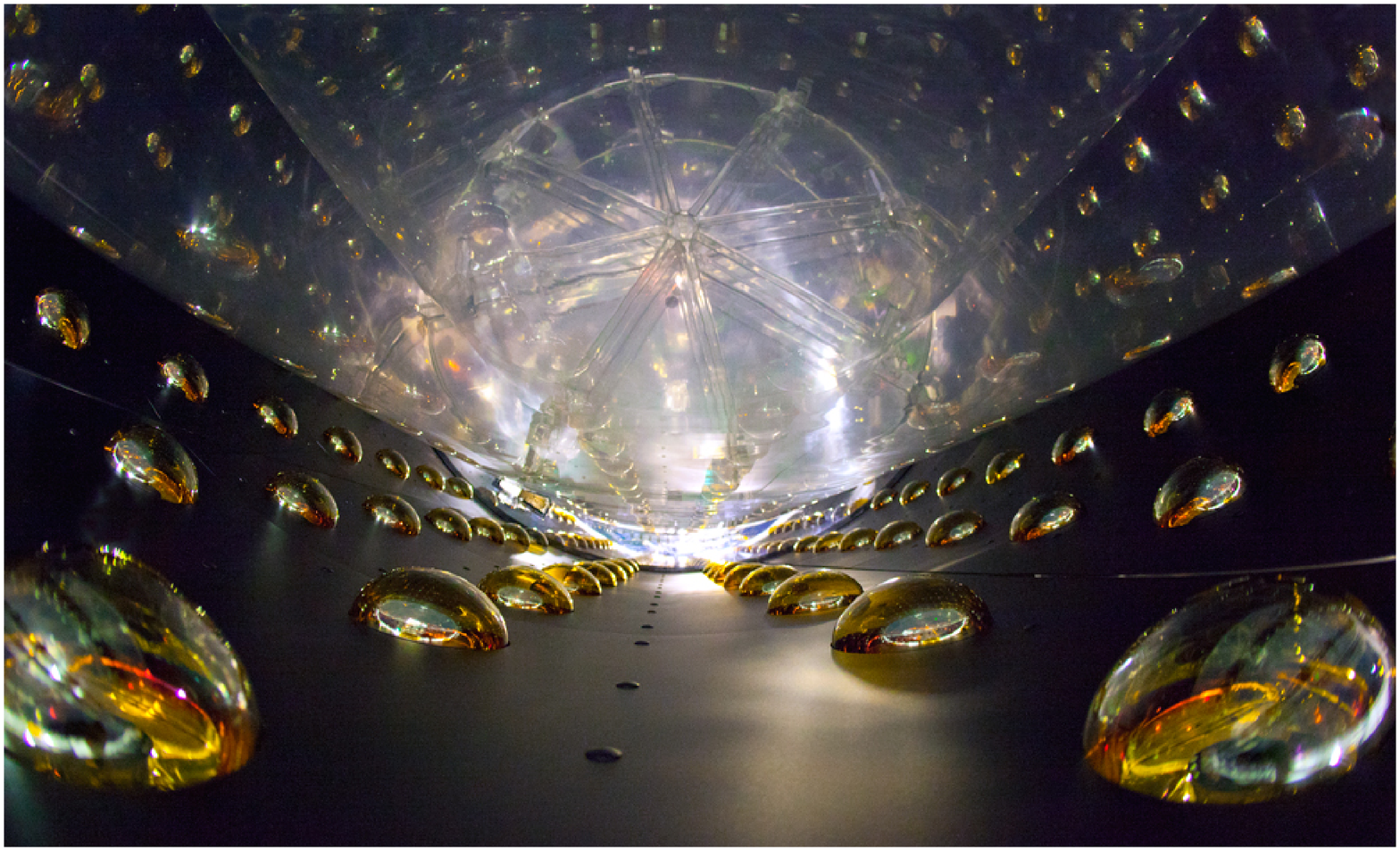}
\caption{The inside of a fully populated AD before top reflector and lid were put on.}
\label{fig:populated-AD}
\end{figure}

The anti-neutrino detector is a sealed stainless steel cylinder. It has a three-layer internal structure.
In the center is the Gd-loaded liquid scintillator serving as the target region.
The second layer is pure liquid scintillator to catch the gamma energy leaking from the target region.
Scintillation light is detected by photo multiplier tubes, PMTs, that are installed in the outer layer
which is filled with mineral oil to shield against the radioactivity from the stainless steel vessel, the PMT supporting structure and the PMTs themselves.
192 PMTs are deployed in 8 rings in the mineral oil region, along the cylinder wall, facing inward, to measure the
energy deposited in the AD. A sectional drawing is in Fig.~\ref{fig:AD-sectioned}.
There are no PMTs on the top and bottom, instead highly reflective panels are mounted to increase the
light collection efficiency.
The inside of a fully populated AD is shown in Fig.~\ref{fig:populated-AD} before the AD lid and top reflector were put on.
The two inner acrylic vessels and some PMTs are visible.

Three Auto-Calibration-Units (ACU) are installed on the AD lid. They can be used to scan the detector response
along three axes, the central axis, A, an off-center axis in Gd-loaded target region, B, and
another off-center axis in the pure liquid scintillator region, C.
The optional sources are LED, germanium-68 and americium-carbon with cobalt-60.

The synthesis of 185 tons Gd-loaded liquid scintillator and 180 tons
pure liquid scintillator are finished, which is all needed for three halls.
The Gd-loaded liquid Scintillator, currently stored in 5 40-ton tanks, is
continually circulated and nitrogen purged.
The precision of the target mass is ensured by redundant mass measurements during AD filling,
including a 20-ton calibrated filling tank and a coriolis mass flowmeter.
The uncertainty of the mass measurement is only 4kg out of 20 tons.
Long term monitoring of the optical and scintillation properties shows all
liquid sample are stable.

The first two ADs have finished construction and liquid filling and
are already locked down to their positions in the Daya Bay Near Hall (see Fig.~\ref{fig:Filled_EH1}).

\subsection{Muon Veto}
The signal IBD event consists of a prompt and a delayed signal.
The energy range of the paired signals and the coincidence window are so special that
muons by themselves can hardly make a background.
But the muon-induced spallation background, neutrons and other long-lifetime heavy nuclei
can be very serious. An active muon veto detector is still needed to tag these muons.
During offline study the coincidence of muon and muon-induced background will
be reconstructed and their contribution will be calculated.

\begin{figure}
\centering
\includegraphics[width=120mm]{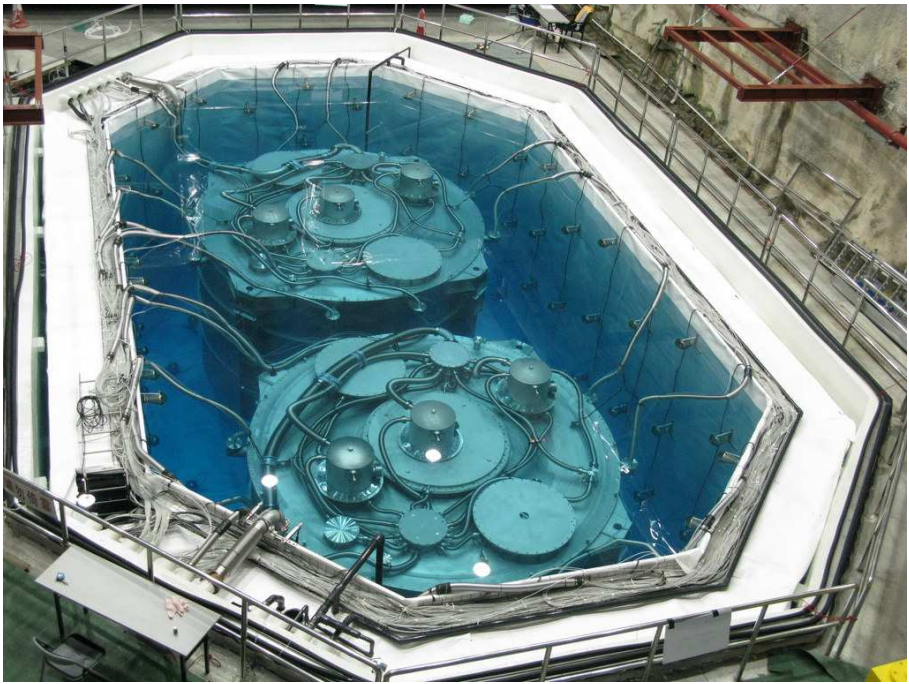}
\caption{Filled Daya Bay near hall water pool with two ADs installed.}
\label{fig:Filled_EH1}
\end{figure}

As previously mentioned the muon veto includes two parts, the water pool and RPCs.
PMTs are installed in the water pool to detect the cerenkov light generated by cosmic ray muons in water.
The water pool itself is separated into two regions, inner and outer, by reflective tyvek sheets.
This increases the light collection efficiency and provides a redundancy in muon detection.
Fig.~\ref{fig:Filled_EH1} shows the status of the Daya Bay Near Hall:
Two ADs are installed. Water is filled to its nominal level.
The inner water pool is visible while outer pool is completely blocked by white tyvek sheets.
RPC modules are assembled and installed in parallel with water pool construction. The RPC modules will
cover the whole top surface of the water pool to provide better position information for part of
the down-going muons.

\subsection{Electronics}
All PMT electronics for the Daya Bay Near Hall running are installed and ready to use.
The PMT front end electronics can provide a timing accuracy of 1.5625ns.
The shaping time for charge measurement is 300ns. Several types of
triggers can be issued determined by multiplicity, number of fired PMTs within a time window,
analog energy sum, external signals etc. All on-site clocks are
synchronized to a GPS signal, so the Daya Bay Experiment can serve as a
supernova observatory.
The RPC electronics are still under commissioning and will come online soon.

\section{AD Dry Run}
Before the filling with liquid, all PMTs and other instruments in the ADs were tested
with the full data processing chain. Raw PMT signals were processed by electronics, and the
charge and time of each PMT hit were recorded if a trigger command was issued.
Data files were automatically transferred to two major computing sites simultaneously,
Lawrence Berkeley National Laboratory in the US and the Institute of High Energy Physics in China,
with a latency of only a few minutes. Data quality and detector performance were monitored automatically.
The ADs and electronics show good results with these intensive tests.

\begin{figure}[h]
\centering
\includegraphics[width=80mm]{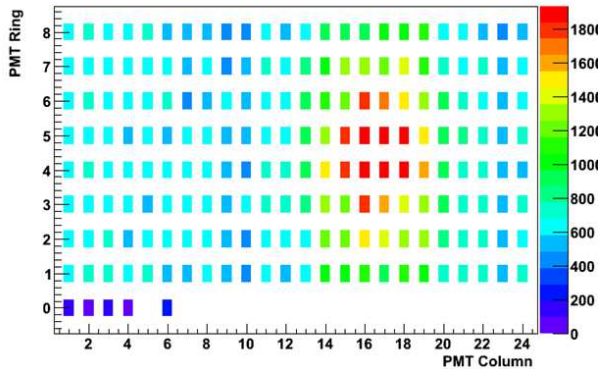}
\caption{192-PMT charge distribution in ring and column for off-center LED source at half height of AD. }
\label{fig:ACUB-LED}
\end{figure}

\begin{figure}
\centering
\includegraphics[width=80mm]{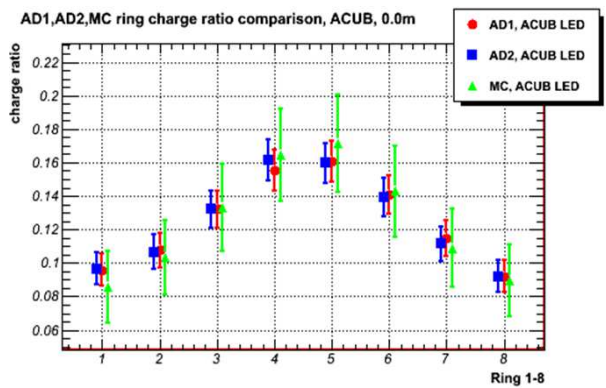}
\caption{AD No. 1, 2 and Monte-Carlo charge ratio comparison.}
\label{fig:Charge-ratio}
\end{figure}

\begin{figure}
\centering
\includegraphics[width=80mm]{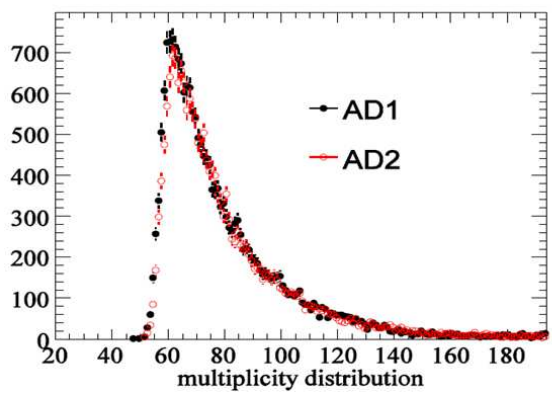}
\caption{AD No. 1 and 2 multiplicity distribution comparison.}
\label{fig:Multi-Comp}
\end{figure}

Some initial offline analyses were done to further understand the
properties of the ADs. In the following the dry run results for AD No. 1 and 2 will be shown.
A LED enclosed by a plastic diffuser ball was lowered into AD to study the response of PMTs.
Fig.~\ref{fig:ACUB-LED} shows the result when one off-center LED (ACUB) was flashing at the half
height of the detector.
What shown is the charge distribution in ring and column (a projection of an unrolled AD),
and each spot is color-coded according to the total charge of a PMT in that run.
Ring 0 is for 2-inch liquid monitor PMTs.
Light is more concentrated around ring 4, 5 and column 17 where PMTs are closer to the LED diffuser ball.
With no liquid filled the charge distribution basically shows the geometric effect between LED
and PMT positions. It is also affected by the optical properties of all kinds of surfaces in the AD, i.e.
the surfaces of two acrylic vessel, top and bottom reflectors, etc.
In Fig.~\ref{fig:Charge-ratio} the charge ratio in each ring
(total charge in one ring divided by total charge in whole detector),
is plotted for the same LED configuration. Results of AD No. 1 and 2 and simulation are
overlaid together, where good consistency can be observed.
Another indication of the identicalness of AD No. 1 and 2 is shown in Fig.~\ref{fig:Multi-Comp}.
When a high energy muon flies through the thin acrylic wall of the two inner acrylic vessels,
cerekov light is emitted which is bright enough to trigger the detector.
Since AD No. 1 and 2 were tested side by side, it is believed that the cosmic ray muon
within two detectors should have identical fluxes and momentum spectra.
In Fig.~\ref{fig:Multi-Comp} the multiplicity of these muon events of AD No. 1 and 2
are plotted. They are again identical within uncertainty.
Note that the cut-off around 50 is caused by trigger threshold.

\begin{figure}
\centering
\includegraphics[width=80mm]{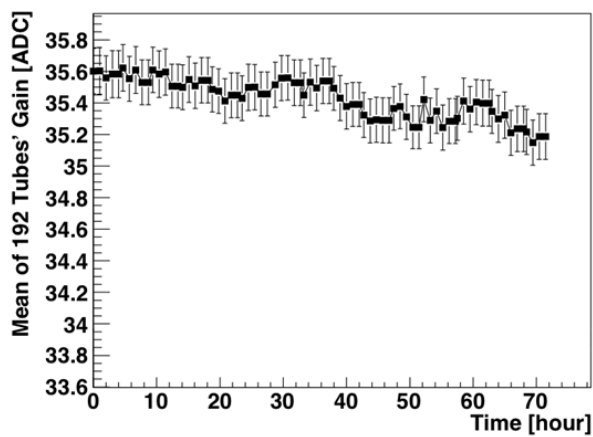}
\caption{Average gain of 192 PMTs in a stability run.}
\label{fig:gain-72h}
\end{figure}

Besides all kinds of functionality tests for the ADs, a long stability run was also arranged.
All PMTs, high voltage supply, front end electronics, trigger system and data acquisition system
were run for over 72 hours. For example, in Fig.~\ref{fig:gain-72h} shows the average gain of
192 PMTs in one AD within this testing period where PMT gain is presented in units of ADC which
describes the final digital output of the PMT signal after electronic shaping.
The variance of the average PMTs' gain is less than 1\%.

\section{Summary and Outlook}
To achieve the challenging goal of measuring $\st13$ to a precision 0.01 at 90\% C.L.,
great efforts were put into the design and construction of the Daya Bay Neutrino Experiment.
The analysis of AD No. 1 and 2 dry run data shows the good status of the detectors.
The beginning of data taking with two ADs in the Daya Bay Near Hall
is anticipated in the summer of 2011 and data taking with all eight
ADs in three halls in the summer of 2012.

\begin{acknowledgments}
It is a pleasure to thank the organizers of "DPF 2011" for the interesting conference
at Brown University. I am also very grateful for all the people who contributed to the
Daya Bay Neutrino Experiment. At the same time I'd like to thank Laurence Littenberg,
Steven Kettell, David Jaffe and David Webber for discussing the contents of this talk and
a careful reading of the manuscript.
\end{acknowledgments}

\bigskip 

\end{document}